\documentclass[aps,prb,onecolumn,groupedaddress,showpacs]{revtex4}
\usepackage[english]{babel}
\usepackage{epsfig}
\usepackage{graphicx}% Include figure files

\usepackage{dcolumn}% Align table columns on decimal point

\usepackage{bm}% bold math

\begin{document}
\title{ Property trends in simple metals: an empirical potential approach} 
\author{A.~Nichol and G.J.~Ackland}
        
\affiliation{School of Physics, SUPA and CSEC, The University of Edinburgh, Mayfield Road, Edinburgh,
EH9 3JZ, United Kingdom}

\date{\today}

\begin{abstract}
We demonstrate that the melting points and other thermodynamic
quantities of the alkali metals can be calculated based on static
crystalline properties.  To do this we derive analytic interatomic
potentials for the alkali metals fitted precisely to cohesive and
vacancy energies, elastic moduli, lattice parameter and crystal
stability.  These potentials are then used to calculate melting points
by simulating the equilibration of solid and liquid samples in thermal
contact at ambient pressure. With the exception of lithium, remarkably
good agreement is found with experimental values. The instability of
the bcc structure in Li and Na at low temperatures is also reproduced,
and, unusually, is not due to a soft T1N phonon mode. No forces or finite
temperature properties are included in the fit, so this demonstrates a
surprisingly high level of intrinsic transferrability in the simple
potentials. Currently, there are few potentials available for the alkali metals, so in addition to demonstrating trends in
behaviour, we expect that the potentials will be of broad general use.
\end{abstract}


\maketitle

\section{Introduction}

The thermodynamic properties of Group 1A metals vary systematically
down the group.  Cohesive energies and elastic constants decrease from
Li-Cs, while lattice parameters increase. This makes them an ideal
testground for testing thermodynamic relationships between solid
properties and melting points.

Melting points are impossible to calculate analytically, and it is
well known that different parameterizations of metal potentials can
give very different results for calculated melting points.  Thus to
examine trends down a group, we require an analytic description of 
interatomic interactions.  This means creating an interatomic potential 
fully specified by a small number of materials properties. 

We wish to derive a family of interatomic potentials describing the
alkali metals which can be used in classical molecular dynamics.  To
ensure comparability, we aim for a form with a minimal number of
parameters, fully determined by the fitting data. We choose to use the
simplest form which describes many-body metallic
interactions, the second-moment approximation to tight
binding\cite{ducastelle1970moments,finnis1984simple}. 
The motivation for this theory comes from the idea of a local density of states projected onto an atom, but the actual potentials are similar in form to  
the embedded atom method\cite{daw1984embedded} (EAM).  This is based on the
conceptual idea of embedding an atom into a preexisting charge
density, and calculating the energy change.

In either case the energy is written as:

\begin{equation}
\mathcal{U} = \sum_i{F({\rho}_i)} + \frac{1}{2}\sum_{i,j\neq i}{{V}_{ij}(r_{ij})}
\end{equation}

with 

\begin{equation}
{\rho}_i= \sum_{j\neq i}{\phi_{ij}(r_{ij})}
\end{equation}

For elements where binding comes from a
single band, the second-moment approach to second moment potentials
gives $F=\sqrt{}$, although for materials where binding comes from two
bands this is more complicated\cite{ackland2003two,ackland2006two}.
For alloys, another subtle difference arises. In EAM  $\phi_{ij}$ is
interpreted as the charge density due to atom $j$, such that
$\phi_{ij}= \phi_{j} \ne\phi_{ji}$, while in the tight binding picture
it represents a hopping integral, and $\phi_{ij}=\phi_{ji}$.

The second moment approximation does not account for the shape or
filling of the band, this is implicit in the parameterization.
Consequently, to examine trends in behavior, we should consider
materials with similar band shapes and band fillings. It appears that
the alkali metals provide an ideal case where this will work well:
they have simple half-filled $s$-band binding and minimal $s-d$ and
$s-p$ hybridization.

Parameterization of potential models can follow two paths: maximal or
minimum fitting parameters.  In the first case, one tries to achieve
the most highly tuned potential by fitting to as many known properties
as possible.  This gives the best possible description of a particular
material. In the second, used here, one uses a minimal set of fitting
data.  Such potentials may not reproduce materials properties as
successfully, but if they do for a whole group of materials, it
demonstrates transferrability due to the physics, and a physical
connection between the fitted and unfitted properties.  Furthermore,
simple potentials should be treated as a null hypothesis, and a
systematic {\it failure} to predict properties indicates missing
physics, even if the properties can be reproduced by judicious
fitting.  For example, we see later that neglect of zero-point effects 
in Lithium increases the calculated melting point.

Our main aim is to ensure comparability between the metals, rather
than transferability of a particular potential.  Consequently, we
consciously eschew approaches such as MEAM\cite{baskes1992modified},
two-band\cite{ackland2003two} and REAM\cite{zhou2013response} which
have added complexity which gives the possibility to fit more closely
to experimental data.  We also avoid overconstrained data fitting
which would allow tuning to a particular
element\cite{brommer2007potfit}.  Without doubt, additional fitting
parameters could be used to ``tune'' melting points, phase
transitions, or other properties of interest, but our interest here is
the intrinsic transferability of the potentials. 

\section{Calculations}
\subsection{Alkali Metals and Periodicity}
\label{SEC:AlkaliMetals}
The alkali metals comprise the group I elements excluding
Hydrogen. They are particularly soft metals with low melting points
and all adopt a body-centered cubic (bcc)
crystal structure at standard temperature and pressure. Their bcc
lattice parameters are notably large, resulting in low densities and
high compressibilities. Being group I elements they have a [noble gas]
+ ns\textsuperscript{1} electronic structure and have been studied
extensively as a test of theories of `simple' metals.

The solid/liquid phase transition in various metals has been studied
using both classical and quantum mechanical methods and with both
Monte Carlo and Molecular Dynamics approaches.  Many-body potentials
for one-off alkali metals have been created
before\cite{wilson2015solid,vella2014comparison,belashchenko2012embedded,ouyang1994calculations},
but they have problems with crystal stability and have not yet been
applied by other groups.  Typical discrepancies between experimental
and simulated melting points are on the order of \emph{hundreds} of
degrees.\cite{mendelev2007development}

The functions $\phi$ and $V$ were defined as cubic splines $F$ is given by
its Finnis-Sinclair\cite{finnis1984simple}  form ($F = {\rho}^{1/2}$) and
\begin{equation}
V(r) =  \sum_{k=1}^6{a_k{(r_{k}-r)}^3}H(r_{k}-r)
\end{equation}
\begin{equation}
\phi(r) =  \sum_{k=1}^2{A_k{(R_{k}-r)}^3}H(R_{k}-r)
\end{equation}
where $H(x)$ is the Heaviside step function: $H(x)=0$ for $x<0$ and $H(x)=1$ for $x>0$. 

The spline knot points ($r_k$ and $R_k$) are not treated as adjustable parameters, their values
were taken from previous work\cite{han2003interatomic}, scaled by the
lattice parameter to give a potential extending to second neighbours.
The 7 parameters $a_k$
and $A_k$ were chosen to exactly fit target values for cohesive
energy, lattice constant, three elastic constants, unrelaxed vacancy
formation energy and fcc-bcc energy difference (Table 1).  These properties are
independent of the final parameter $a_6$, which controls the short
ranged repulsion inside the perfect crystal nearest neighbour
distance.  This was set to a constant value across all potentials,
scaled by the cubed lattice parameter.  The $r_k$ parameters were set
to be the same for all elements, as a fraction of lattice parameter 
(see Table 2).

It can be noticed that all energies and elastic moduli systematically
decrease down the period, by factors of about 2 and 6
respectively. However, the reduction in the elastic moduli is almost
entirely due to the increased lattice parameter: the elastic moduli
in units of eV per unit cell volume are remarkably constant.

\begin{table}
\begin{center}
\begin{tabular}{|c|c|c|c|c|c|c|c|}

\hline
Element & $a_0$ & C$_{11}$ & C$_{12}$ & C$_{44}$ & $ E_{coh} $ & $E_{vac}$ & $E_{fcc}$ \\
\hline
\hline
Li   &  3.51  &0.092 & 0.078 & 0.067 & 1.648  &0.54 & -0.006 \\
Na   &  4.2906 & 0.0512 & 0.0418 & 0.0345 & 1.109 & 0.35 & -0.003 \\      
K    &  5.328 & 0.0260 & 0.0213 & 0.0179 & 0.923 & 0.308 & +0.001 \\
Rb   &  5.585 & 0.0213 & 0.0179 & 0.0138  &0.840 & 0.28 & +0.012 \\
Cs   &  6.141 & 0.0162 & 0.0135 & 0.00999 & 0.788 & 0.263 & +0.016\\
\hline
\end{tabular}
\end{center}
\caption{\label{tab:data} Fitting Data: bcc lattice parameter (\AA),
  elastic constants (eV/\AA$^3$) cohesive energy, unrelaxed vacancy
  formation energy and fcc energy per atom (eV) above the bcc value.
  For sodium the low temperature ground state is actually fcc, and for
  lithium it is reported to be a 9R complex close-packed structure,
  however bcc is stable at room temperature, and we fit to the bcc
  properties}
\end{table}

The square root dependence appears to make the fitting nonlinear,
however this nonlinearity can be transferred from the fitting to the
data being fitted. Specifically, the many-body term's parameters
$A_k$ can be fully determined by linear fit to two quantities which
are explicitly zero for any pair potential, i.e.
\begin{itemize}
\item
the difference between vacancy formation energy and cohesive energy $(E_{coh}-E_{vac})$
\item the Cauchy Pressure  $C_{12}-C_{44}$
\end{itemize}
The pair-potential parameters $a_i$ can then be determined by a linear
fit to the difference between the required property and the
contribution from the many-body term.  The lattice parameter is fitted
by setting the derivative of the energy to zero at the required value.
The fitting problem is then reduced to a simple 7x7 matrix problem.

We also attempted the fit using a genetic
algorithm\cite{dieterich2010ogolem}, which converged immediately for
the linearly transformed problem, but failed to find the known
solution within required several weeks of CPU time when fitting the
seven pieces of data directly.  The algorithm got ``stuck'' in many
local minima, all of which are eliminated by transforming the fitting problem
to linear algebra.

 For meaningful comparison, it is useful that the linear fit for all
 potentials gives a broadly similar solution: parameters are given in
 Table \ref{TAB:Parametera}.  The best way to visualize these
 potentials is the ``effective pair potential'', which incorporates both
 attractive and repulsive parts, approximating $\rho$ by its
 equilibrium value at the minimum bcc energy $\rho_0$.

\begin{equation}
V_{eff}(r) =  V(r) + \phi(r)/\rho_0 
\end{equation}

\begin{figure}
\centering
\includegraphics[width=0.9\textwidth]{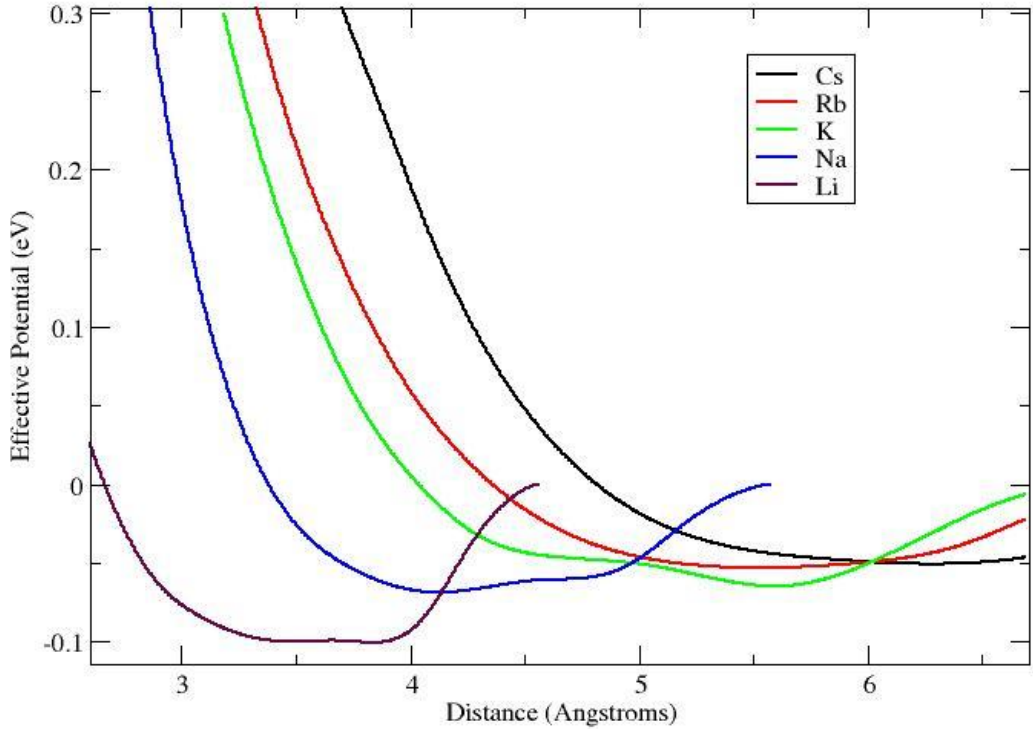}
\caption{Effective pairwise potentials (table \ref{TAB:ResultsTable})
  showing trends down the group.
\label{FIG:AllpotentialsClose}
}
\end{figure}

 These functions $V_{eff}(r)$ shown in
 Fig. \ref{FIG:AllpotentialsClose}.  Several trends are notable here,
 the most obvious being the range and depth of the potentials.  More
 subtle is the development of a secondary minimum in Li, Na and K.
 Although the parameterization data comes from bcc only, this feature
 is ultimately responsible for the stability of fcc in Li and Na at
 low temperature, and K at high pressure.  The existence of such a
 minimum is not explicit in the second moment approach.  If it
 appeared for a single material, we might have dismissed it as a fitting
 anomaly.  However, it is tempting to speculate that it arises from a
 mismatch between the optimal electron density and the first minimum in
 the screened electrostatic potential.  This latter emerges from the
 free electron theory, the oscillations ultimately being related to the
 Fermi wavevector\cite{animaluHeine}.

\begin{table}[htbp]
  \centering
    \begin{tabular}{|l|cccccc|}
\hline
Li &&&&&&\\
$a_k$ &  -9.741070  &  52.683696  & -75.033831  &  39.542714 &  -14.782577  & 3.079201\\
$A_k$  &  20.537818  & -34.233934 &&&&\\
$l.p.$ &    3.510000  &&&&&\\
\hline
Na &&&&&&\\
$a_k$    & -4.194182  &  27.427457  & -44.559331  &  31.771667 &  -15.792378  &  3.079201\\
 $A_k$      & 8.719427 &  -13.899855&&&&\\
$l.p.$ &       4.290600 &&&&&  \\
\hline
K &&&&&&\\
$a_k$ &    -1.723040 &   5.203775 &   4.914958 & -21.814488 &  20.037488  
 &   3.079201\\
$A_k$  &    5.438419    & -8.619811 &&&&\\
$l.p.$ &      5.328000  &&&&&  \\
\hline
Rb &&&&&&\\
$a_k$ &     -2.924102 &   15.101598  & -19.036044  &   8.221297  &   1.000245  &  3.079201\\
$A_k$  &     4.151321  &  -6.163082 &&&&\\
$l.p.$ &      5.585000  &&&&&\\
\hline
Cs &&&&&&\\
$a_k$ &     -2.184268  &   9.803950  &  -8.445145   & -2.009154   &  8.266863  &  3.079201\\
$A_k$  &     3.415288  &  -4.866591 &&&&\\
$l.p.$ &      6.141000  &&&&&\\
\hline
$r_k$ & 1.3 & 1.22 & 1.15 & 1.06 & 0.95 & $\sqrt{3/4}$ \\
$R_k$ & 1.3 & 1.2 &&&&\\
\hline

    \end{tabular}%
  \caption{Tabulated parameters for the alkali metal potentials.  For
    ease of comparison, values are scaled by the lattice parameter
    (l.p.), such that the units are eV per lattice parameter cubed.
    This enables up to use consistent values for $R_i$ and $r_i$,
    enhancing comparability of the potentials \label{TAB:Parametera}}
\end{table}%

\begin{table}[htbp]
  \centering
    \begin{tabular}{|llcccc|}
    \hline
    \textbf{Element} &  & {$\mathbf{T_m^{exp}\,/\,K}$} & {$\mathbf{T_m^{sim}\,/\,K}$} & {$\mathbf{\Delta\,h\,/\,kJ\,mol^{-1}}$} & {$\mathbf{\Delta\,v\,/\,mL\,mol^{-1}}$} \\
    \hline    \hline
    \textbf{Lithium} & Li    & 454   & 660(2) & 44.38(1) & 0.0301(2) \\
%    \textbf{Lithium*} &       &       & 551(2) & 19.30(1) & -0.0223(1) \\
    \textbf{Sodium} & Na    & 370   & 411(2) & 3.51(2) & 0.223(3) \\
    \textbf{Potassium} & K     & 336   & 344(2) & 2.99(2) & 0.211(1) \\
    \textbf{Rubidium} & Rb    & 312   & 333(2) & 2.70(2) & 1.385(1) \\
    \textbf{Caesium} & Cs    & 301   & 305(2) & 2.60(2) & 1.957(1) \\
    \hline
    \end{tabular}%
  \caption{Tabulated results for the alkali metal potentials. Shown are the experimental and simulated melting points, as well as the specific enthalpy and volume differences between the solid and liquid phases at the coexistence temperature.}% The asterisk refers to the modified lithium potential.}
  \label{TAB:ResultsTable}%
\end{table}%
\subsection{Simulating Phase Transitions}
To calculate the melting point, we employ the coexistence
method\cite{coexist} in which pre-formed samples of the two phases are
brought together and allowed to equilibrate. A solid/liquid interface
is present at the start of the simulation and the particle velocities
are initialized to approximate the anticipated coexistence
temperature. The total enthalpy of the system is held constant, in an
NPE ensemble, so that if the initial temperature is below (above)
$T_m$ the phase boundary will move and some of the sample will freeze
(melt).  The ``ringing mode'' is eliminated in an initial
equilibration phase by setting the first time derivatives of the Nose
and Parrinello-Rahman extended Lagrangian parameters to zero whenever
their sign differs from the second derivative.  This algorithm does
not correspond to any Lagrangian, but very efficiently removes energy
from ringing modes.  After a short equilibration period, when the
energy in the fictitious dynamic modes approaches kT, we continue
production runs with the standard NPE Lagrangian.  This process proved
necessary because of the poor coupling between the fictitious degrees
of freedom and the rest of the system.  We note that the same ringing
behaviour is present in the NVE ensemble, manifesting itself as an
oscillation in the internal pressure.
     
\begin{figure}[H]
\centering
\includegraphics[width=\textwidth]{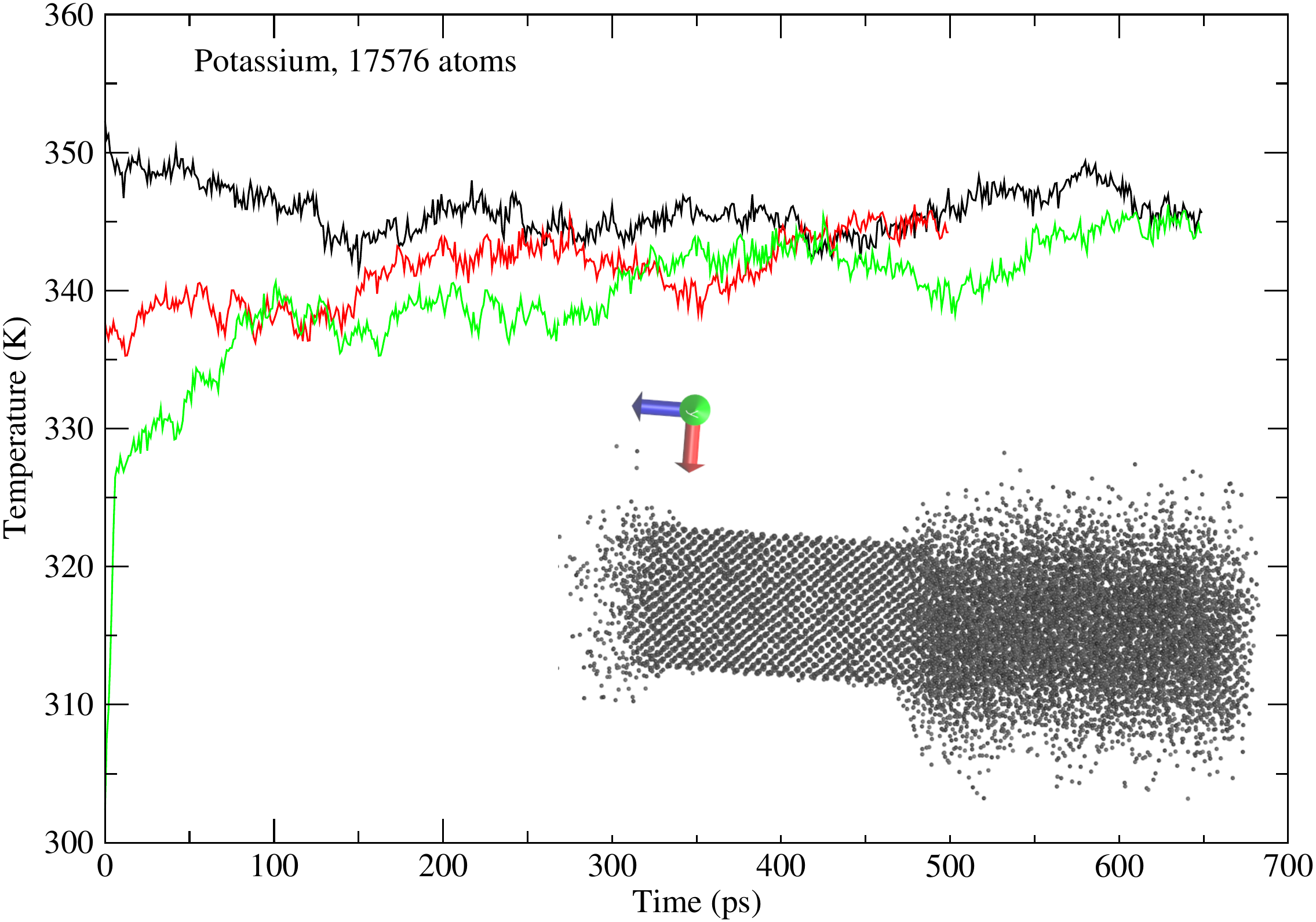}
\caption{Temperature evolution of coexisting crystalline and liquid K
  in the NPEp ensemble. Simulations had temperatures initialized above
  or below. After relaxation all systems equilibrated to a single
  coexistence temperature. Inset is a snapshot of the simulation, with
  the periodic boundary conditions removed in the graphics to show
  the two distinct phases.  Similar behaviour was observed for other elements}
\label{FIG:Kcoexist}
\end{figure}

On a timescale much longer than the
ringing mode equilibration, the latent heat released during the
transition increases (decreases) the temperature of the sample until
it reaches $T_m$.  We used a supercell with 17576 atoms (initialized
with 8788 atoms of bcc: $13 \times 13 \times 26$, and a similar amount
of melt).  Convergence was checked by comparing several runs
initialising the temperature above and below the expected melt point,
and observing that in each case the simulation went to the same final
temperature.  The final configurations were visualized using VMD
\cite{VMD} to verify that both crystal and solid phases were still
present (Fig \ref{FIG:Kcoexist}, inset).

All calculations reported here were carried out using the MOLDY
molecular dynamics package\cite{moldy}.  The potentials were also
converted to LAMMPS format\cite{lammps} and are available at the NIST
potential database\cite{NIST}.  These versions have adjustments at
very short range to fit to the Biersack-Ziegler\cite{BZ} functions which are
popular in radiation damage studies, but give identical results for
the problems considered here.
%Independent tests of the melting point were carried out using coexistence on LAMMPS and Phase Switch Monte Carlo.

\begin{figure}[H]
\includegraphics[width=0.5\textwidth]{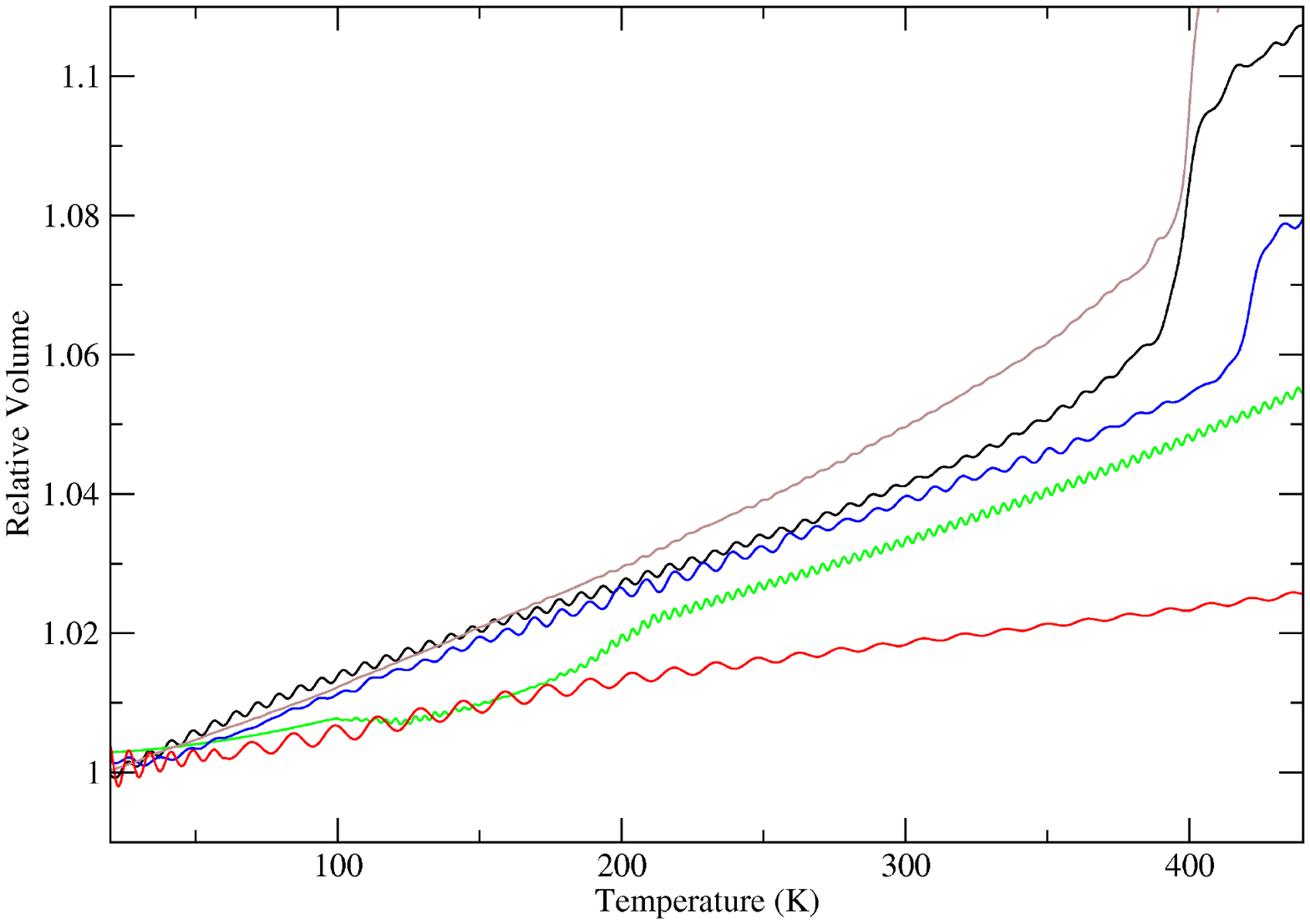}
\caption{Volume against temperature for steady heating in NPT
  molecular dynamics.  Volumes are scaled to their low-T bcc value to
  show trends in thermal expansion across the group, from lowest to
  highest Li (red) Na (green) K (blue) Rb (black) and Cs (brown).
  Heating rate was 1 K per 4000 timesteps, the timestep varying with
  the atomic mass from 0.1fs for Li to 1fs for Cs. Ringing-mode volume
  oscillations come from the Nose thermostat and Parrinello-Rahman
  barostat and persist for hundreds of picosecond. Sharp rises in
  volume around 400K for Cs, Rb, K, indicate melting, although
  hysteresis means this is not the thermodynamic melting point.  The
  large oscillations for Li show the immediate transformation to
  close-packing at low T.  The two changes in slope for Na come from
  the (metastable)bcc-fcc transition at about 100K, and the subsequent
  fcc-bcc retransformation around 200K.
\label{FIG:Thermal}}
\end{figure}

Figure \ref{FIG:Kcoexist} shows the changes in temperature for three
constant-energy simulations of a potassium supercell. 
Convergence times depend primarily on the atomic
mass and on the system size and were typically on the order of hundreds of
picoseconds.
The thermal expansion of the materials was also calculated by constant
pressure heating simulations (Fig \ref{FIG:Thermal}).  Thermal
expansion depends on third derivatives of the potentials, which are not
fitted, nevertheless, both the values and trends across the group are
in good agreement with experiment.  There is a weak temperature
dependence, and in the absence of quantization of the phonons, calculated thermal expansions remain
finite to 0K.  The heating calculations also showed melting
transitions in all elements, and a martensitic transformation to a
faulted close-packed structure at low-T in Li and Na, also in accord
with experiment.  We observe that although this is an excellent test of
phase stability, ``heat until it transforms'' is not a reliable way to determine transition
temperatures.

\subsection{Phase Coexistence Lines at pressure}
\begin{figure}[H]
\includegraphics[scale=0.2]{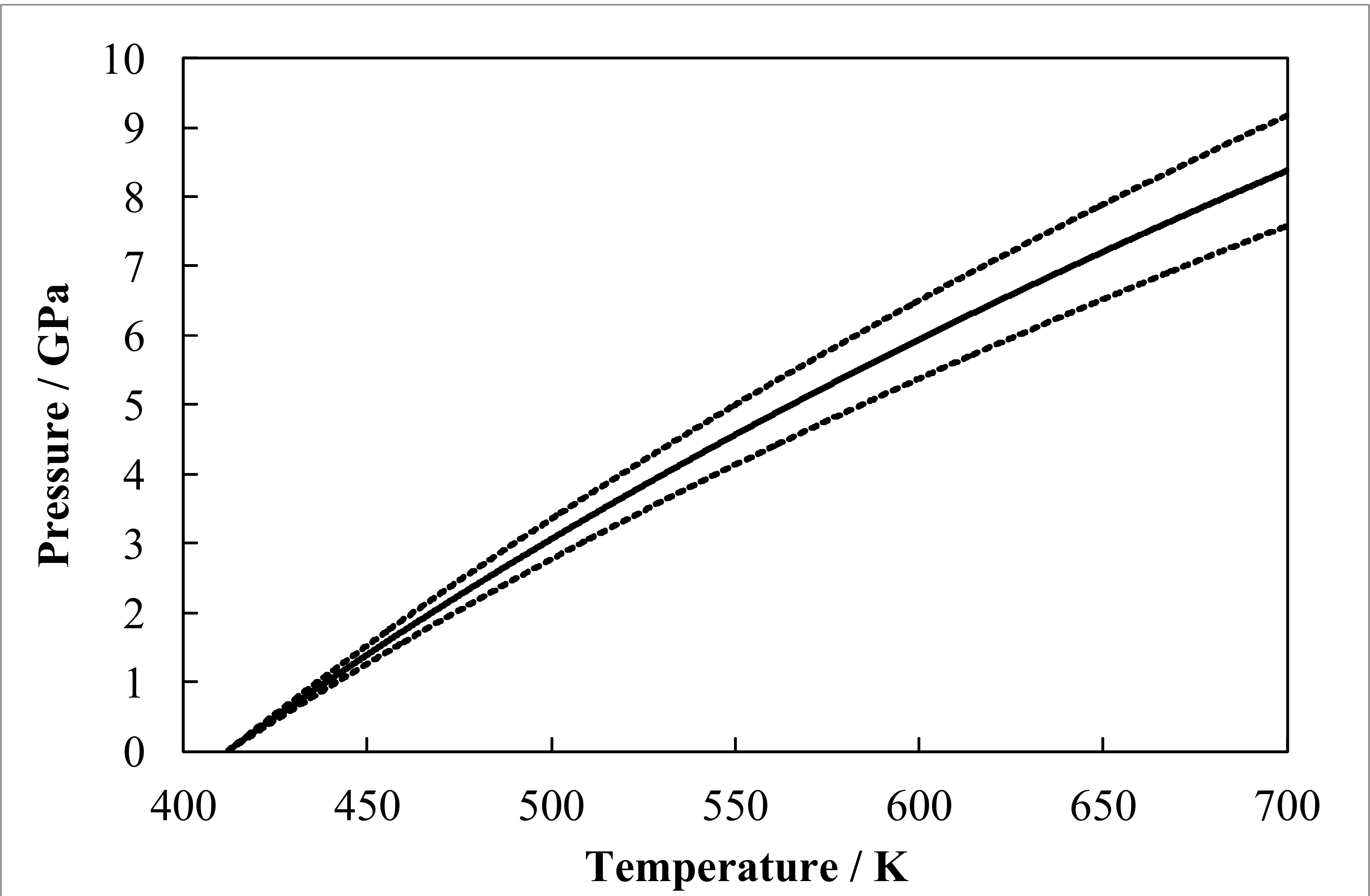}
\centering
\caption{Coexistence line for Na integrated from a single simulation at $P=0.0$ using the Clausius-Clapeyron equation. \textbf{Solid Line}: Best estimate based on the specific enthalpy and volume differences between the solid and liquid at the zero-pressure melting point of 405 K. \textbf{Dashed Lines}: 67\% confidence intervals based on the relative errors in the initial slope.}
\label{FIG:NaPTCurve}
\end{figure}

In order to generate coexistence lines on the PT phase diagram of the
alkali metals the Clausius-Clapeyron equation was integrated using a
fourth order predictor-corrector algorithm. The coexistence conditions
($P$,$T$) were extrapolated over a range of several hundred degrees in
intervals of 0.01 K. Shown in Figure \ref{FIG:NaPTCurve} are the
results for sodium, others being similar.  
%This is compared to explicit calculations using coexistence at different pressures.

At high pressure, the alkali metals are known to exhibit a maximum in
the melting curve.  It is striking that none of the potentials
presented here exhibits this feature: this tells us that physics
beyond that sampled in the ambient pressure state, where the fitting
was done, will be required.  Another high pressure phenomenon,
transformation to fcc, was observed in all the potentials, as a
consequence of the lower bulk modulus and higher packing density under
pressure.  In Na, and Li, there is a martensitic transition to a
complex close-packed structure at low temperatures\cite{LiNa9R,Sodiumfcc}

\subsection{Phonons}
\begin{figure}[H]
\includegraphics[width=0.7\textwidth]{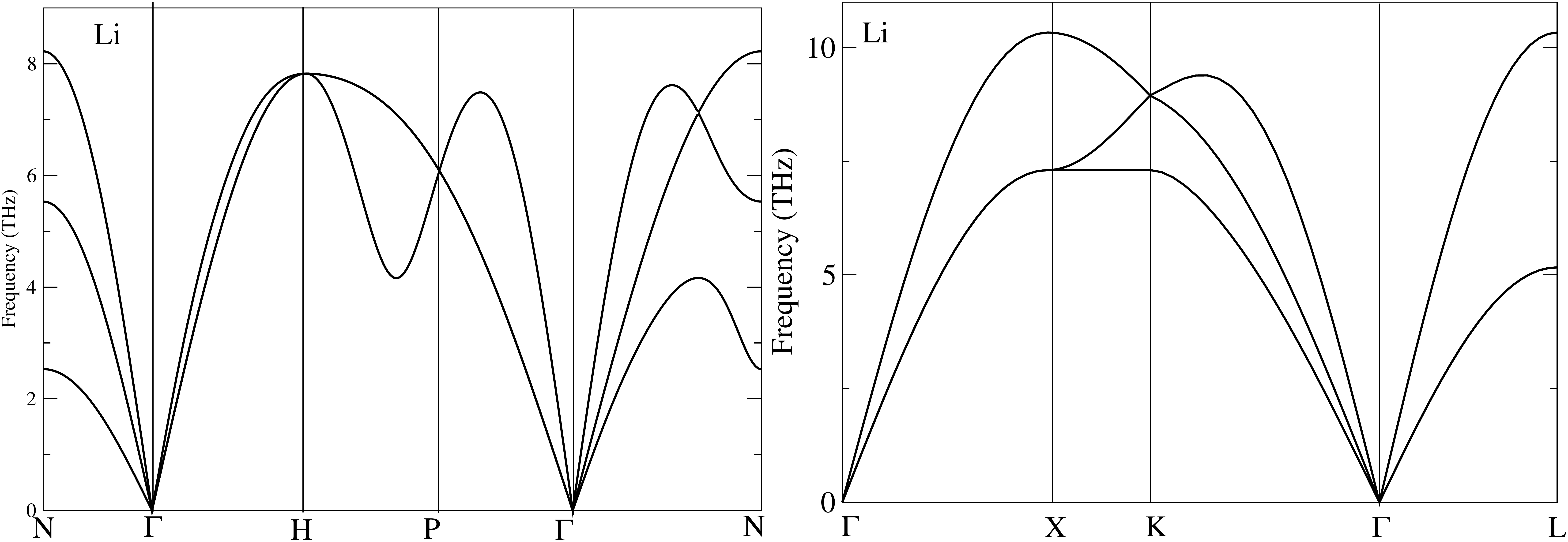}
\caption{Phonon spectra calculated using lattice dynamics in
  lithium in bcc (left, fitted) and fcc (right).
\label{FIG:Phonons}}
\end{figure}

\begin{figure}[H]
\includegraphics[width=0.7\textwidth]{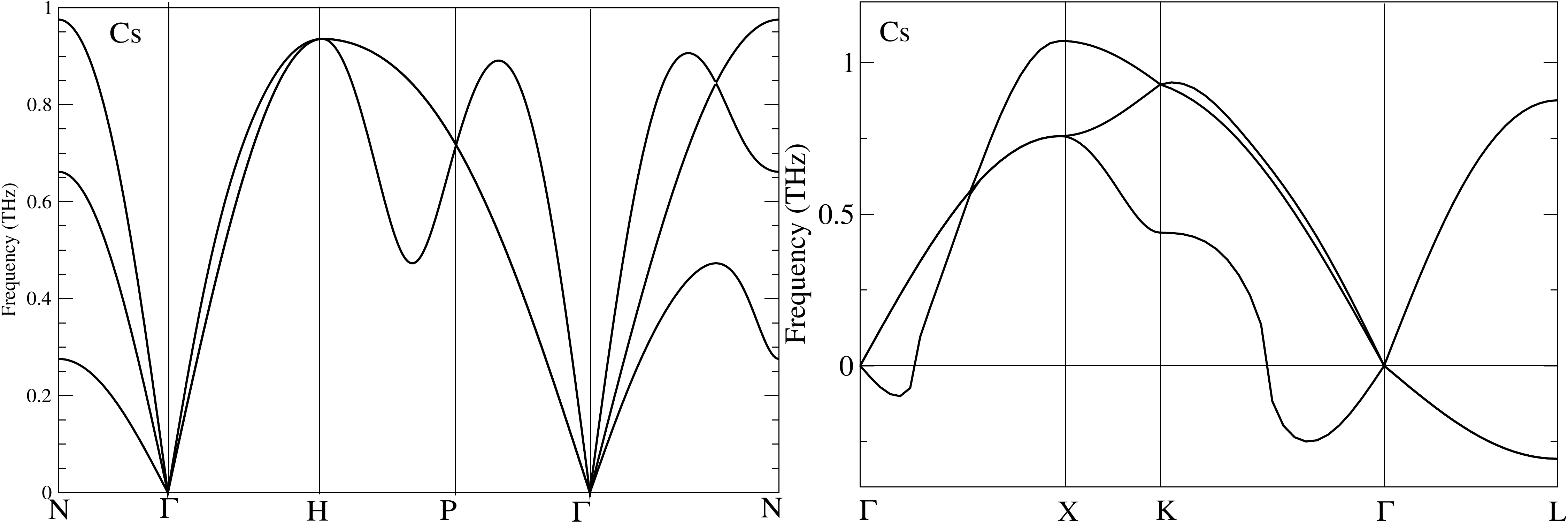}
\caption{Phonon spectra calculated using lattice dynamics in Cs bcc
  (left, as fitted) and fcc (unstable phase).
\label{FIG:PhononsCs}}
\end{figure}

Phonon spectra can be calculated by lattice dynamics using the
analytic second derivatives of the potential.  For bcc, the symmetry
constraints and fit to elastic constants mean that agreement with
experiment is exact for the slope at the $\Gamma$ point, so the
general similarity with neutron data is unsurprising and trends across
the group can be readily seen.  It is notable that Li and Na, which
are unstable with respect to fcc at low temperature, have stable bcc
phonon spectra, unlike other elements such as Ti and Zr, which are
calculated to have negative bcc shear moduli at 0K\cite{Tegner,may1999,mendelev2007development}, and
transform to bcc at high temperatures due to soft T1N modes. Figures
\ref{FIG:Phonons} and \ref{FIG:PhononsCs} show phonon spectra for Li
and Cs, other elements being similar (Supplemental Figure 3).  The
stable phases have real phonons throughout, but notably, fcc Cs is
mechanically unstable with imaginary phonon modes along $\Gamma-L$.
At higher pressures when Cs becomes thermodynamically stable, all
phonons are stable.

It is also noticeable that the phonon frequencies are higher in fcc
than in bcc.  This has the effect of stabilizing the bcc phase at
higher temperatures.  It is notable that this feature is present in
all cases, despite the properties of fcc elements being excluded from
the fitting process.

Coexistence lines for the remaining alkali metals are shown in supplemental materials.

\subsection{Parameter sensitivity}
Our fitting process is fully defined, however there are uncertainties
in the quantities being fitted.  We tested the melting points against
variation of the fitting parameters for lithium, and it proved to be
remarkably robust except with respect to the short-range part of the
repulsion (inside 3.03975$\AA$, parameter $a_6$), which is
uncorrelated with the fitting data and chosen to match smoothly to the
Biersack-Ziegler\cite{BZ} functions.  Stiffer short range repulsion
accords to higher melting point (and thermal expansion).  We speculate that this is because the steep repulsion
reduces the available phase space for the liquid state by more than
the solid, reducing its relative entropic advantage. It is
possible to lower the melting point by up to 100K before other
unreasonable behaviour appears - notably that the liquid density
becomes lower than the solid and the thermal expansivity becomes
negative. 

The very good agreement elsewhere suggests that for lithium our model
is missing some physics.  A reasonable candidate is the nuclear
quantum effects, which are especially important for low mass effects.
The classical molecular dynamics does not include the zero-point
energy (ZPE) of lithium, which we calculated in our lattice dynamics
and by using the CASTEP density functional package\cite{CASTEP} to be
around 39meV/atom, corresponding to a Debye temperature of around
450K, somewhat higher than a previous study\cite{moruzzi}.  ZPE
dominates the vibrational contribution to the free energy below the
Debye Temperature, which in lithium is close to the melting point.
The effect on the melting point is that the quantum vibrational
entropy is about half the classical value, and must therefore be
assumed to be significant.  The solid lithium ZPE is also
significantly larger than the latent heat of melting (13meV/atom),
which indicates that the liquid ZPE is also significant.  

An accurate assessment of nuclear quantum effects on the melting point
would require evaluating the ZPE for the liquid, for which there is no
suitable theory.  Qualitatively, we can expect that the zero point
energy of the solid will be greater than that of the liquid, since the
liquid has no resistance to shear. Consequently, nuclear quantum
effects will destabilize the solid compared to the liquid, lowering
the melting point.  This is consistent with our observation that our 
calculated classical melting point is too high.

\section{Discussion and Conclusions}

We have derived a series of analytic many-body interatomic potentials
which describe the simple metals.  Although the potentials are fitted
only to crystal properties, they give a good description of the
melting point with the notable exception of lithium, which we ascribe
to nuclear quantum effects ignored in our MD.

The potentials are fitted to fcc and bcc energy at 0K, and to
properties of bcc at ambient conditions, Consequently, the transition
into a higher temperature bcc state in Li and Rb is a successful
prediction.  It is due to the lower phonon frequencies in bcc: the
only contributory factor which was fitted is the bcc elasticity
tensor.  Heuristically, the fcc stability can be attributed to the
appearance of a secondary minimum in the effective potential, which is
more pronounced in the lower Z-elements.  This minimum is not an input 
to the theory, it emerges from
the fitting process, implying that its existence manifests in some
feature of the fitted bcc elasticity.
 
Overall, the work shows that a number of trends can be deduced from
minimal fitted data.  Essentially, the input is that energies and
elastic moduli are lower for higher-Z elements, while lattice
parameters are larger.  From this we deduce that the thermal expansion
and volume change on melting increases down the group, while the
melting point is reduced.

\subsection{Acknowledgements}
We thank The University of Edinburgh for the CPU time allocation,
EPSRC for support under EP/F010680/1.  Open data for the potentials in
this paper is available at http://www.ctcms.nist.gov/potentials/ \\

\bibliography{Refs}
\end{document}